\documentclass[oldversion]{aa} 
\usepackage{graphicx}
\usepackage{aalongtable}
\usepackage{txfonts}
\usepackage{rotating}
\usepackage{lscape}
\newcommand{\gsim}{\;\lower.6ex\hbox{$\sim$}\kern-7.75pt\raise.65ex\hbox{$>$}\;}
\newcommand{\lsim}{\;\lower.6ex\hbox{$\sim$}\kern-7.75pt\raise.65ex\hbox{$<$}\;}

\begin{document}

\title{Multiple stellar populations in the high-temperature regime: Potassium 
abundances in the globular cluster M~54 (NGC~6715)\thanks{Based on observations collected at 
ESO telescopes under programmes 081.D-0286 and 095.D-0539.} 
 }

\author{
Eugenio Carretta\inst{1}
}

\authorrunning{Eugenio Carretta}
\titlerunning{Abundances of potassium in M~54}

\offprints{E. Carretta, eugenio.carretta@inaf.it}

\institute{
INAF-Osservatorio di Astrofisica e Scienza dello Spazio di Bologna, Via Gobetti
 93/3, I-40129 Bologna, Italy}

\date{}

\abstract{Among the multiple stellar populations in globular clusters (GCs) the
very high-temperature H-burning regime, able to produce elements up to
potassium, is still poorly explored. Here we present the first abundance
analysis of K in 42 giants of NGC~6715 (M~54) with homogeneous abundances of
light elements previously derived in our FLAMES survey. Owing to the large mass
and low metallicity, a large excess of K could be expected in this GC, which is located
in the nucleus of the Sagittarius dwarf galaxy. We actually found a spread in
[K/Fe] spanning about 1 dex, with [K/Fe] presenting a significant
anti-correlation with [O/Fe] ratios, regardless of the metallicity component in
M~54. Evidence for a K-Mg anti-correlation also exists, but this is
statistically marginal because of the lack of very Mg-poor stars in this GC. We
found, however, a strong correlation between K and Ca. These observations clearly
show that the K enhancement in M~54 is probably due to the same network of
nuclear reactions generating the phenomenon of multiple stellar populations, at
work in a regime of very high temperature. The comparison with recent results in
$\omega$ Cen is hampered by an unexplained trend with the temperatures for K
abundances from optical spectroscopy, and somewhat by a limited sample size for
infrared APOGEE data. There are few doubts, however, that the two most massive
GCs in the Milky Way host a K-Mg anti-correlation.
}

\maketitle

\section{Introduction}

Multiple stellar populations in globular clusters (GCs) are defined as groups
of  stars that are distinct by differences in light elements (mainly C, N, O,
Na, Mg, Al, and Si) within stellar systems generally showing a very homogeneous
chemical  composition. Since these variations are found at every evolutionary
stage and the currently observed low-mass stars in GCs cannot produce the
species found enhanced in GC stars, the changes must have been imprinted in the
proto-cluster gas by self-enrichment from polluters of an early first generation
(FG; see Gratton et al. 2001). These still unknown actors, being more massive,
went extinct, leaving at least a second  generation (SG) of stars, whose fossil
record is accessible through direct observations.

The different chemistry of multiple stellar populations can be directly studied
through abundance analysis with low- and high-resolution spectroscopy, or
through the effect of varying chemical composition on photometric sequences of
GCs observed in a variety of filters, with bandpasses including features of the
light elements. Despite the efforts of systematic spectroscopic surveys (such as
our FLAMES survey: Carretta et al. 2006, 2009a,b, 2010a and further
developments) and photometric surveys (e.g. Lardo et al. 2011, Monelli et al.
2013, Nardiello et al. 2018), a precise identification of the objects
responsible for early enrichment of GCs is still elusive, and the comparison of
observation and theory is still plagued by unsolved problems (see e.g. the
review by Bastian and Lardo 2018 and Gratton et al. 2019). 

The nature of the involved process is, however, rather well assessed. For a
long time (see the reviews by Smith 1987 and Gratton et al. 2004), the well
defined pattern of anti-correlations and correlations among the light elements
has been recognised as the action of proton-capture reactions in H-burning at high
temperatures (e.g. Denisenkov and Denisenkova 1989). What makes it problematic to
isolate a single candidate FG polluter is the variety of stellar objects able to
reach the necessary temperature for the nuclear burning required to deplete C,
O, and Mg, simultaneously elevating the levels of N, Na, Al, and Si, as observed
in GC stars. Therefore, it is crucial to gather observational material on those
poorly studied species, such as potassium, which are produced only in presence
of very high temperatures (above 180 million K: Iliadis et al. 2016, Prantzos et
al. 2017), together with more extensively investigated species such as Na and
Al.
The simultaneous presence of excesses in K and other species that should have
been destroyed at such high temperatures may exacerbate problems  in the
theoretical models, or even falsify some of the proposed scenarios for early
self-enrichment in GCs.

Vice versa, playing with the different chemical species produced or destroyed in
different conditions, we can hope to pinpoint the temperature (i.e. the mass) 
range allowed for potential polluters. A classical example is the abundance
analysis in \object{NGC 6388} (Carretta and Bragaglia 2019). In this massive bulge GC, we
found that  Mg is not anti-correlated to either Ca or Sc (at odds, e.g., with
NGC~2808). On the other hand, a clear Mg-Si anti-correlation is observed, fixing
a lower limit at 65 million K, which is the temperature required for a
significant leakage from the Mg-Al cycle on Si (Arnould et al. 1999). The
inference that the central temperature of the FG polluters was comprised in a
limited range of values ($\sim 100-150$ MK or $110-120$ MK, depending on the
selected class) then follows.

A systematic census of excesses in Ca and Sc in about a half of the GCs in the
Milky Way was carried out in Carretta and Bragaglia (2021) using unpolluted
field stars as a comparison sample. We detected statistically significant
variations related to proton-capture processes at very high temperatures in
eight GCs out of 77 examined. Ca and Sc are usually well
studied species for many stars in many GCs, but the direct litmus test would be
retrieving correlations and anti-correlations involving K, which unfortunately
is much less investigated. A situation we are helping to somewhat alleviate in
the present paper, which is organised as follows.

After a brief excursus on previous abundance analyses of K abundances in GCs and
a discussion on the potential targets (Sect. 2), we illustrate observations and
analysis (Sect.3), and results (Sect.4) for \object{NGC 6715} (M~54), the
nuclear star cluster associated with the disrupting dwarf galaxy, Sagittarius.
We summarise our findings in Sect. 5.

\section{Target and comparison clusters}

In Carretta et al. (2014, their Fig. 15), we introduced a diagnostic plot to
select candidate GCs where the very high-temperature regime was probably at
work. Using more than 200 stars in several GCs observed at
high resolution with the UVES spectrograph as a reference baseline, we plotted the [Ca/Mg] ratio as a
function of [Ca/H]. The idea was to pick up candidate stars with larger than
average excesses of Ca or depletions in Mg (or a combination of the two). In
particular conditions it is predicted that the consumption of Mg may bypass the
usual production of Al to reach heavier nuclei, starting from nuclei of Ar
(Ventura et al. 2012). The scanty observations that existed at the time confirmed
significant correlations among Ca, Sc, and K abundances in \object{NGC 2419}
(Cohen and Kirby 2012, Mucciarelli et al. 2013, Carretta et al. 2013a),
suggesting that their excesses probably originated from the same source.

\begin{figure*}
\centering
\includegraphics[scale=0.32]{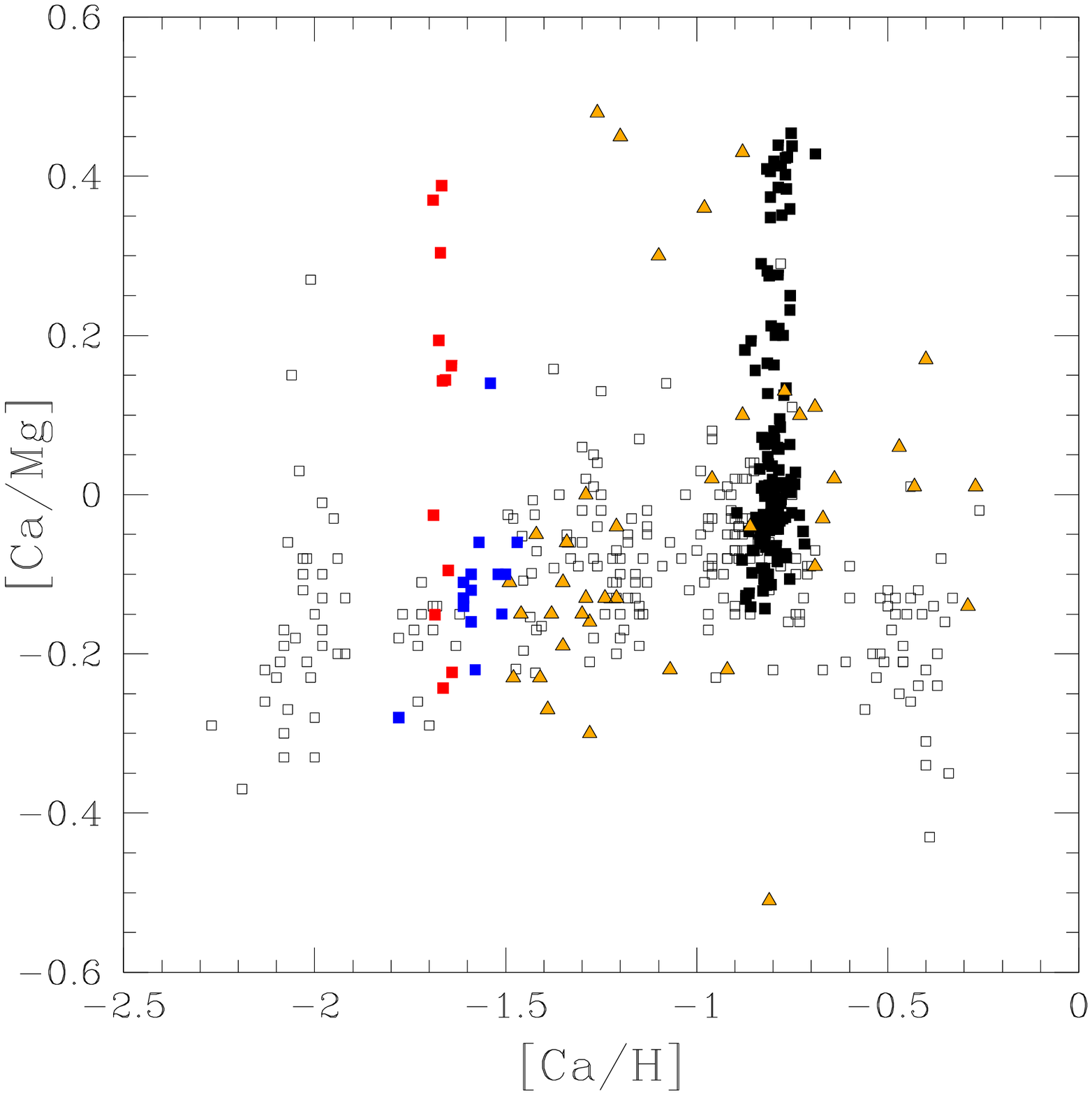}\includegraphics[scale=0.32]{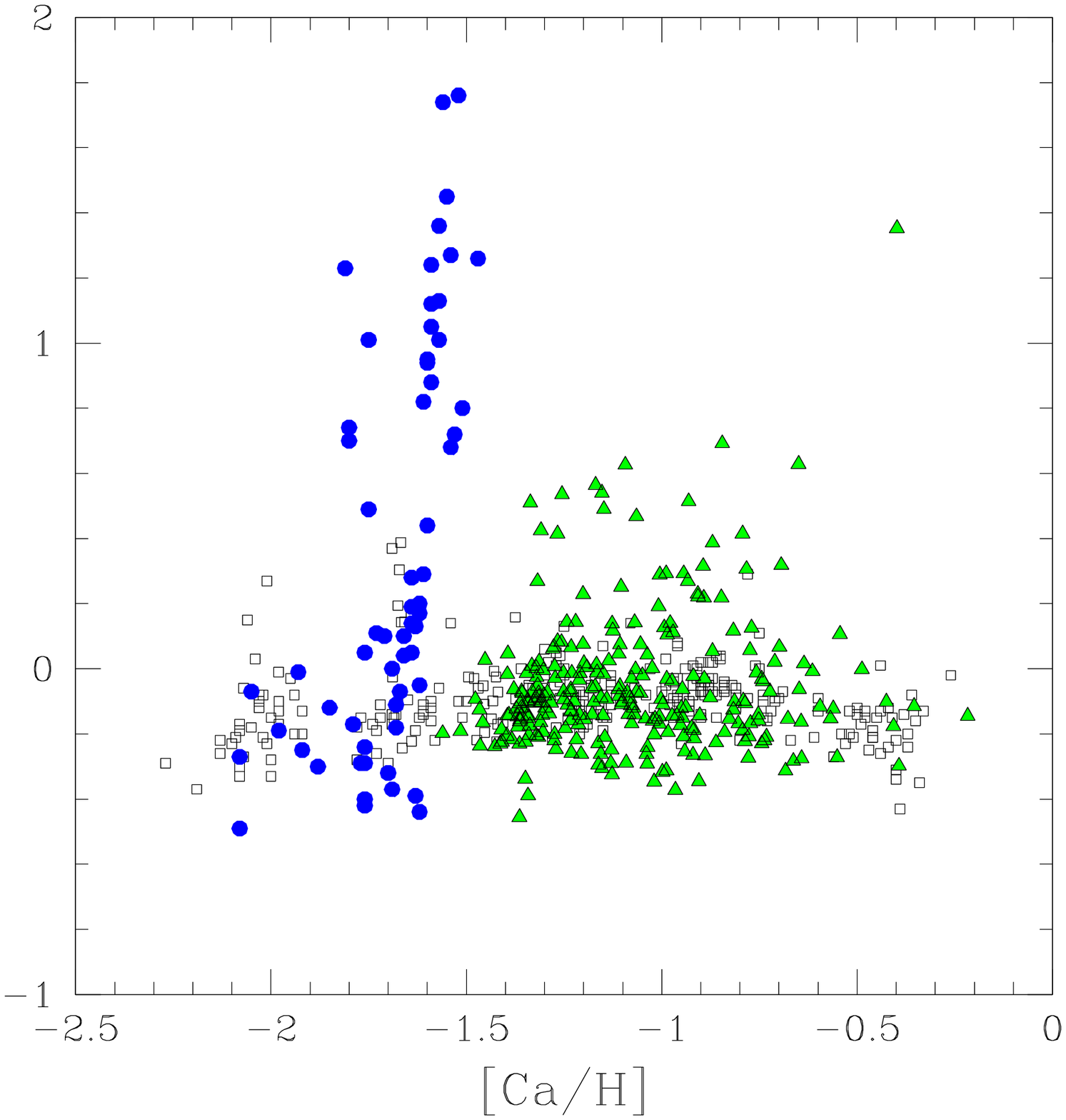}\includegraphics[scale=0.32]{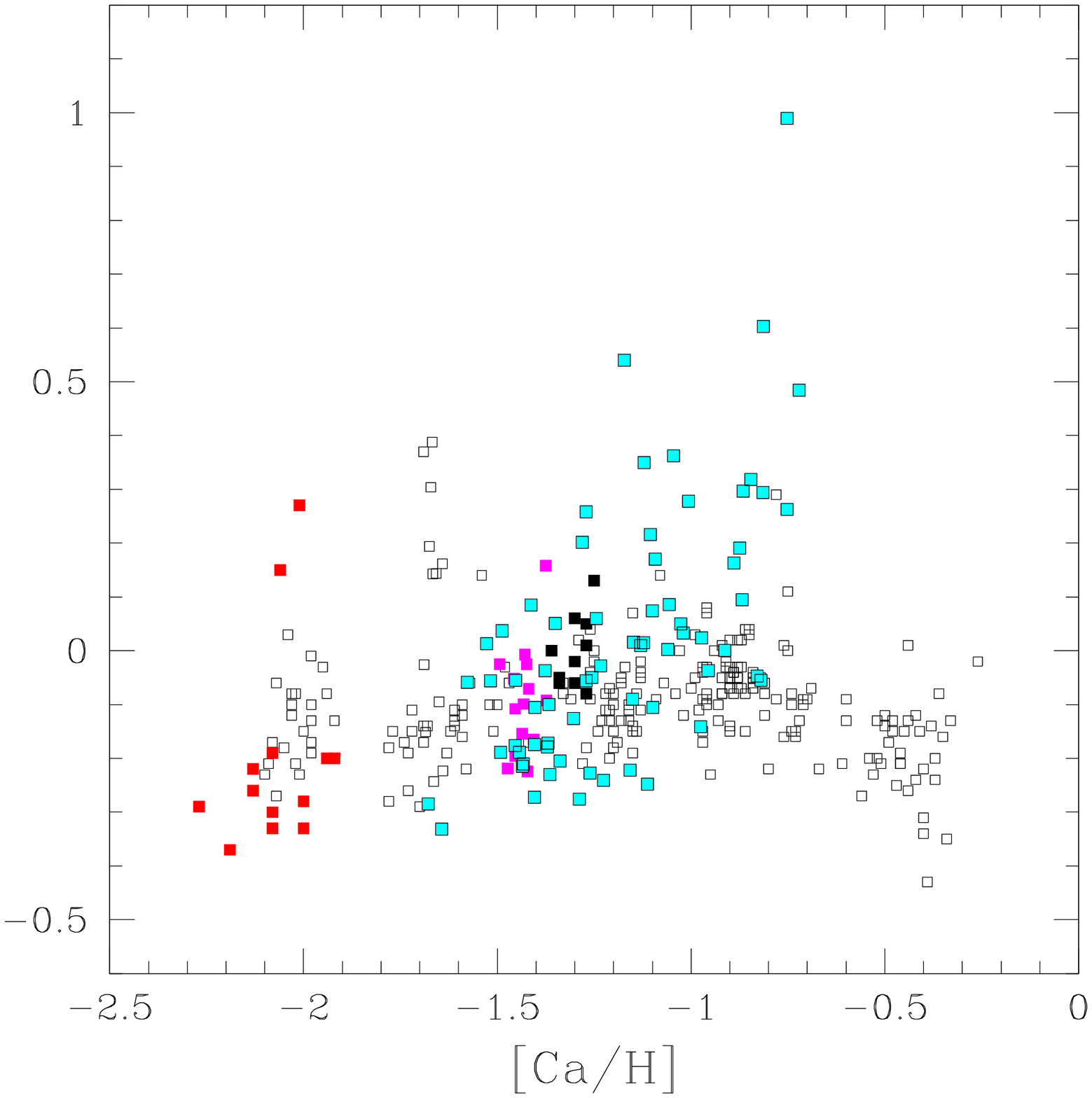}
\caption{Abundance ratios of [Ca/Mg] as a function of [Ca/H] in several GCs. In
all three panels the empty squares indicate stars with UVES spectra in our
FLAMES survey from Carretta et al. (2010b, 2009b, 2010c, 2011, 2013b, 2014,
2015). In each panel a few GCs are highlighted. Left panel: \object{NGC 4833}
(Carretta et al. 2014, red squares), \object{NGC 6809} (Carretta et al. 2010b,
2009b: blue squares), \object{NGC 2808} (Carretta 2015: black squares), and
\object{NGC 5139} ($\omega$ Cen, Norris and Da Costa 1995: orange triangles).
Middle panel: NGC~2419 (Cohen and Kirby 2012, Mucciarelli et al. 2012: blue
circles) and $\omega$ Cen from APOGEE infrared data (M\'esz\'aros et al. 2020,
green triangles). Right panel:  \object{NGC 7078} (M~15: Carretta et al. 2010b,
2009b: red squares), \object{NGC 1904} (Carretta et al. 2010b, 2009b: black
squares), \object{NGC 6093} (M~80: Carretta et a. 2015: magenta squares), and
NGC~6715 (M~54: Carretta et al. 2010c: cyan squares). The different vertical
scale in all panels should be noted.}
\label{f:fig1}
\end{figure*}

An update of that diagnostic plot is provided in Fig.~\ref{f:fig1}. In the left
and central panels, we highlight the GCs scrutinised so far to explore the
high-temperature regime of multiple populations, with particular attention to K
abundances. In the left panel of Fig.~\ref{f:fig1},  NGC~2808 (black squares)
stands out, with high values of the [Ca/Mg] ratios due to the significant
anti-correlation between Ca and Mg stressed by Carretta (2015), together with a
correlation between Ca and Sc. The addition of K (Mucciarelli et al. 2015)
completes the set of heavy proton-capture elements studied in this GC. Red
squares indicate NGC~4833, where K abundances and related correlations were
studied by Roederer and Thomson (2015) and, more extensively, by Carretta
(2021). Blue squares are for NGC~6809 where Mucciarelli et al. (2017) found
marginal evidence for a [K/Fe]-[O/Fe] anti-correlation, compatible with the
solitary star with a high [Ca/Mg] ratio in the left panel of Fig.~\ref{f:fig1}.
Finally, high [Ca/Mg] ratios are seen in stars of $\omega$ Cen (from Norris and
Da Costa 1995, orange triangles), strongly suggesting that the most massive GC
in the Galaxy is a very promising candidate where one can look for large
variations also in K.

This is strengthened by the middle panel in Fig.~\ref{f:fig1}, where we plot 
the [Ca/Mg] ratios in $\omega$ Cen from the APOGEE infrared data  (M\'esz\'aros
et al. 2020, green triangles) together with the abundances in NGC~2419 (Cohen
and Kirby 2012, Mucciarelli et al. 2012). Again, it seems promising to look at
$\omega$ Cen, although the remarkable values shown by NGC~2419 are far to be
reached.

\begin{figure*}
\centering
\includegraphics[scale=0.32]{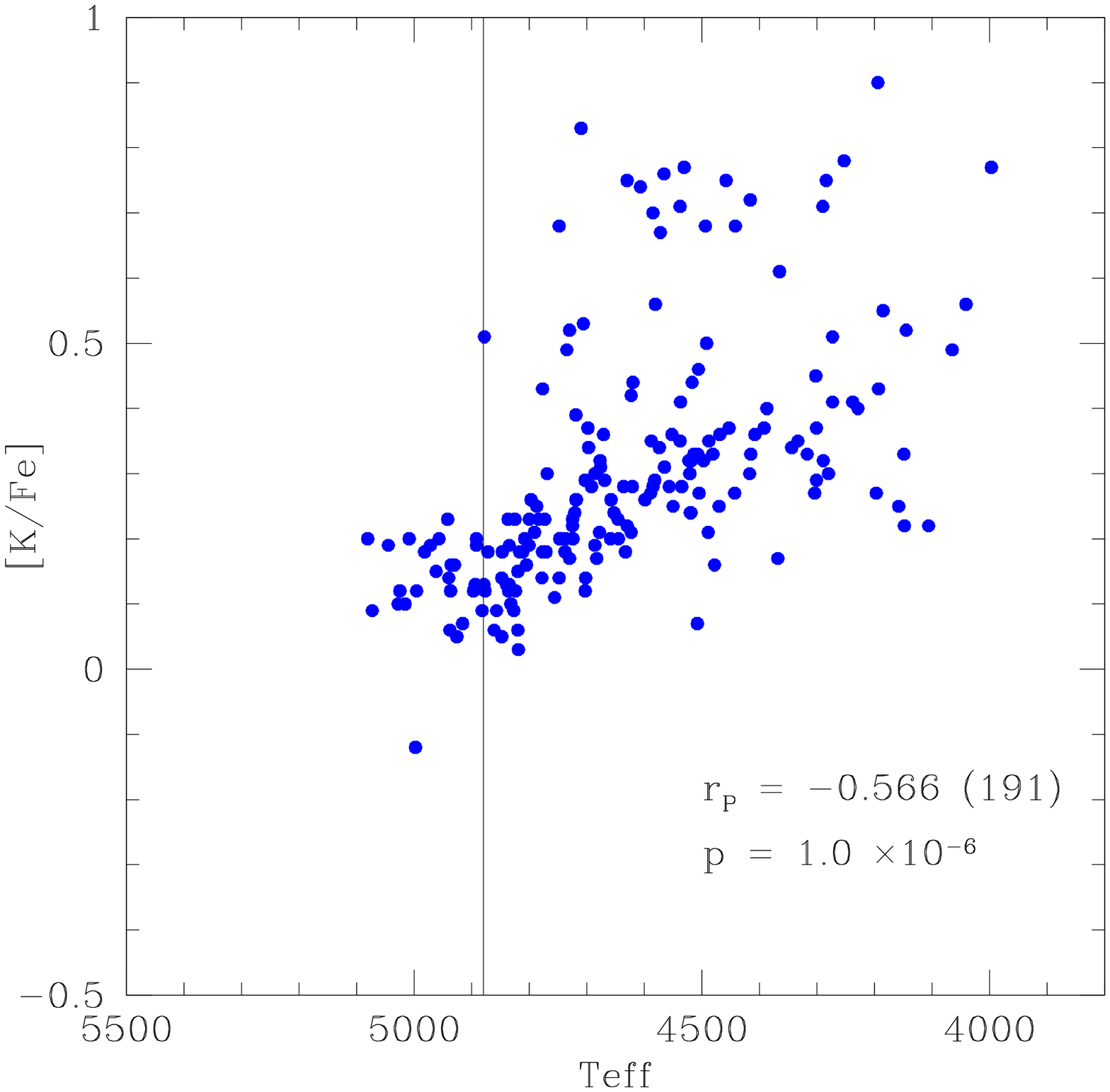}\includegraphics[scale=0.32]{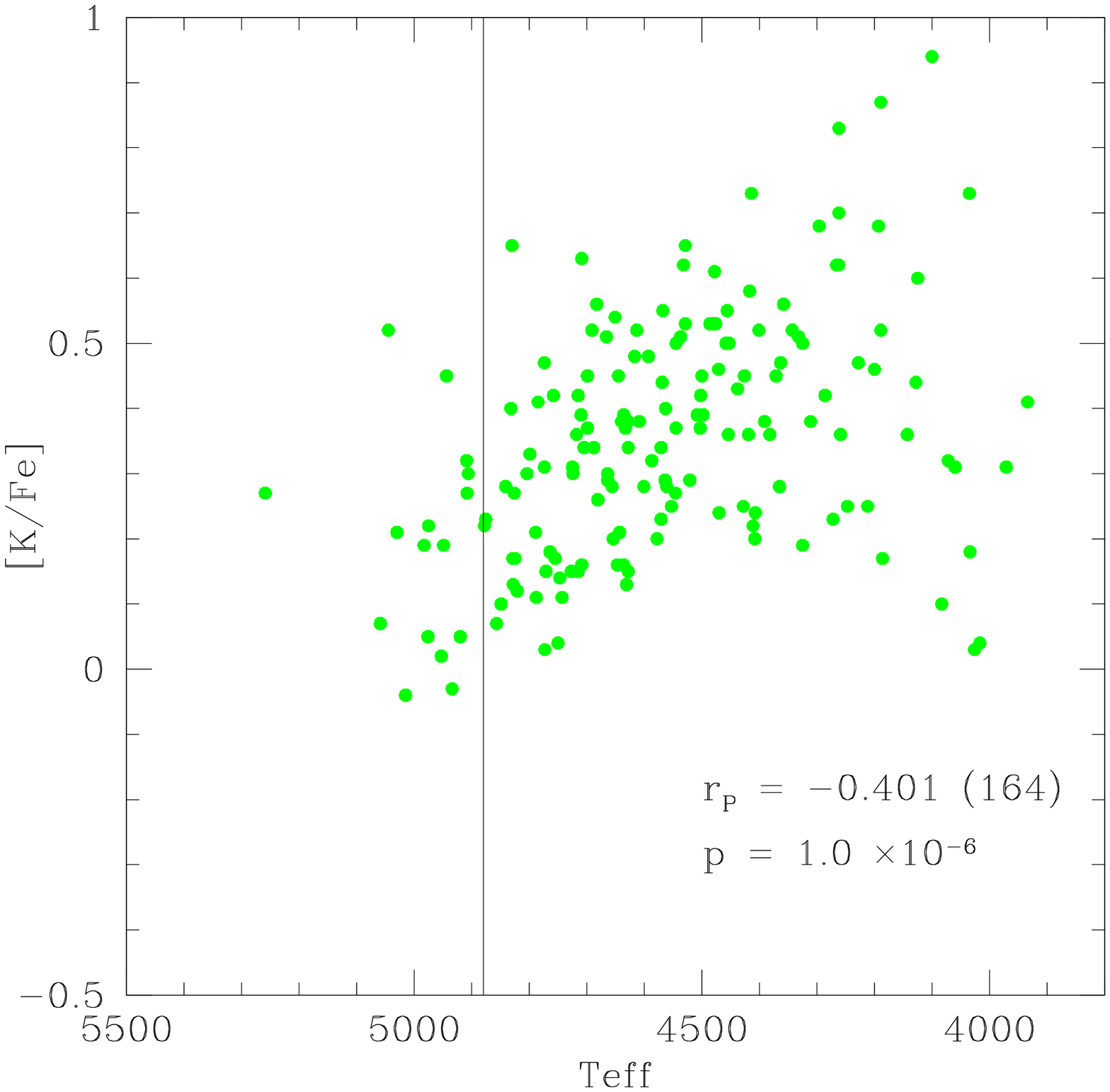}\includegraphics[scale=0.32]{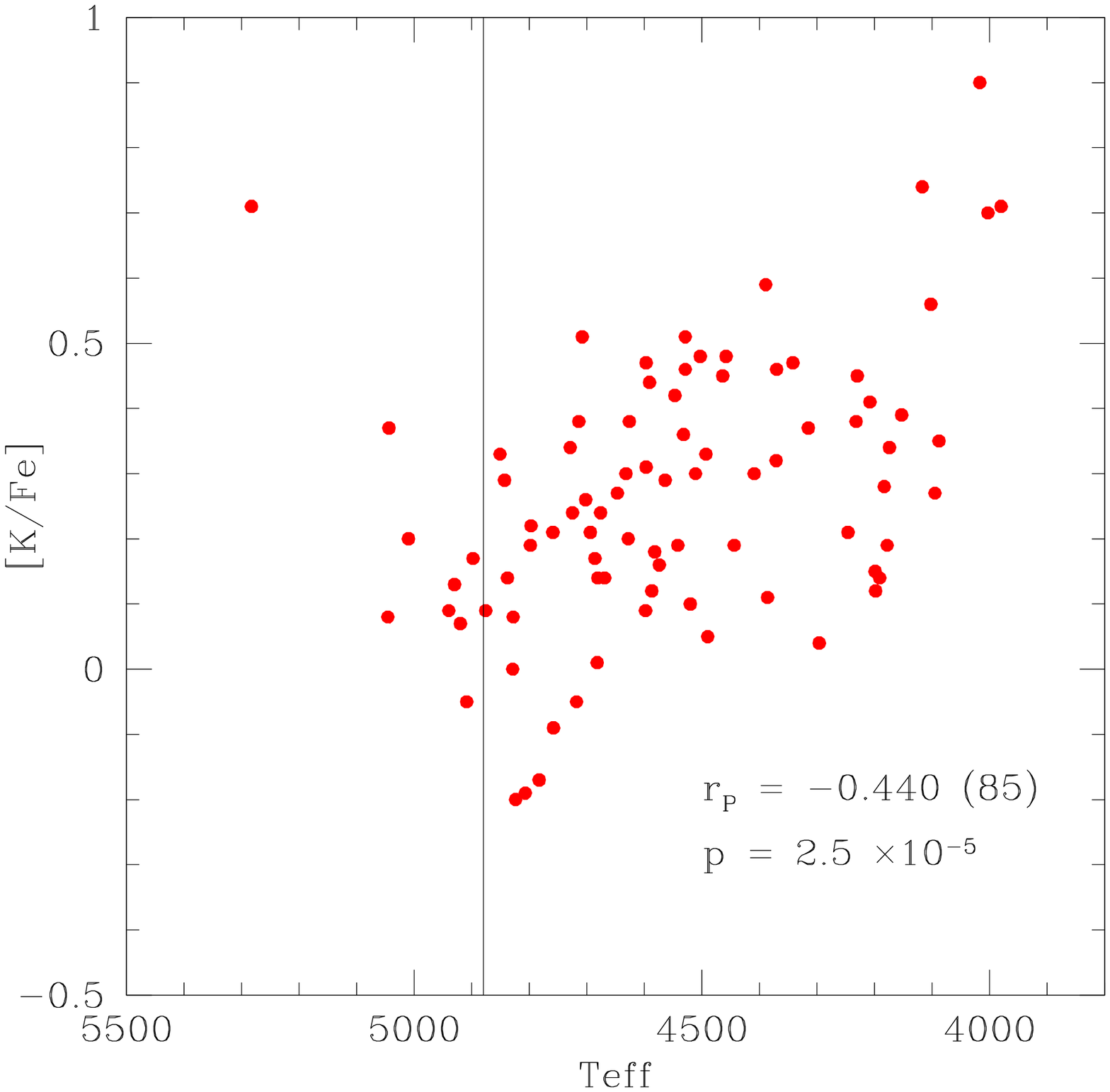}
\caption{Ratios of [K/Fe] as a function of the effective temperature in stars of
$\omega$ Cen from the recent analysis by Alvarez Garay et al.(2022: AG22). Stars
are separated in metallicity bins [Fe/H]$\leq -1.70$ dex,  $-1.70<$[Fe/H]$\leq
-1.30$ dex, and [Fe/H]$>-1.30$ dex, from left to right . The vertical line
indicates the approximate temperature level of the RGB bump in $\omega$ Cen
($G\sim 14.0$ mag) on the temperature scale of AG22. In each panel we report the
Pearson linear regression correlation coefficient, the number of stars in the
metallicity bin, and the two-tail probability of the linear regression occurring
by mere chance.}
\label{f:fig2}
\end{figure*}

A recent analysis of K in $\omega$ Cen from optical spectra (Alvarez Garay et
al. 2022; hereinafter AG22) cannot be used, due to unexpected and unexplained
trends of K abundances as a function of the effective temperature (see
Fig.~\ref{f:fig2}). These trends are all statistically very significant, they
are seen at every metallicity, and they present a monotonic increases in K
abundances as a function of temperature, gravity, and/or magnitude
all the way above the luminosity level of the RGB bump, above which
the stars should show a completely homogeneised envelope, no more affected by
any kind of mixing. Since no such trend is present in the other abundances by
AG22 (Fe, Na, or Mg) we infer some problems in the analysis related to the K,
either in measurements or the application of non-local thermodynamic equilibrium
(NLTE) corrections. 

Finally, in Fig.~\ref{f:fig1} (right panel) GCs still to be explored for K
abundances and showing moderately high [Ca/Mg] ratios are highlighted: M~15 (red
squares), NGC~1904 (black squares) and NGC~6093 (M~80), although these high
ratios are restricted to only a few stars in each GC. To these GCs, we also add a
large sample of stars with GIRAFFE spectra analysed by Carretta et al. (2010c)
in M~54 (NGC~6715: cyan squares). 

After $\omega$ Cen, M~54 is the second most massive GC currently located in  the
Milky Way. These two GCs may even follow a similar path of dynamical evolution,
albeit at different times (see Carretta et al. 2010d). With a clear, intrinsic
metallicity dispersion, M~54 is located in the highest-density region of the
Sagittarius dwarf galaxy, presently in its disruption phase in our Galaxy (Ibata
et al. 1994). 
For the purposes of the present work, the distribution in Fig.~\ref{f:fig1} 
shows that M~54 is the ideal target where to look in order to study the multiple
population phenomenon in the high-temperature regime, by exploiting the largest
set of homogeneous abundances of proton-capture elements derived from 
high-resolution spectroscopy in this GC (Carretta et al. 2010c).

\section{Observations and abundance analysis}

GIRAFFE spectra taken with the high-resolution setup HR18 (R=19,000) were
downloaded from the ESO archive. They were taken under the ESO programme 095.D-0539
(P.I. Mucciarelli), the same as those used for the analysis of similar data in NGC~4833 (Carretta
2021). The one-dimensional and wavelength-calibrated spectra cover the spectral
range from about 7460 to about 7883~\AA, including the resonance doublet of
K~{\sc i} at 7698.98~\AA\ and 7664.91~\AA.

The original observations were made by two exposures of 2600 sec on 19 July
and 19 August 2015. We cross-matched the observed stars with the 76 RGB stars
analysed in M~54 by Carretta et al. (2010c), finding 43 stars in common. The
spectra were sky subtracted. We decided to drop the exposure acquired on 19
August, since the spectra were of lower quality and S/N, to avoid introducing
additional noise. From the remaining set, we disregarded one star with a very low
S/N spectrum. Our final sample consists in 42 giants in
M~54 with  atmospheric parameters and abundances of Fe and light-elements O, Na,
Mg, Si, Ca, and Sc from Carretta et al. (2010c). Coordinates and magnitudes for
the whole sample can be found in Carretta et al. (2010c) in the CDS repository.
All the abundances and the  atmospheric parameters are reported for convenience
in Table~\ref{t:datam54}, together with the S/N of the spectra.

\begin{table*}
\centering
\caption{Atmospheric parameters and abundances of selected stars in M~54}
\begin{tabular}{lllllrrrrrrrr}
\hline
star    &T$_{\rm eff}$ & $\log g$ & $v_t$ & [Fe/H]{\sc i}& [O/Fe] & [Na/Fe]& [Mg/Fe]& [Si/Fe]& [Ca/Fe]& [Sc/Fe]{\sc ii} &[K/Fe] & S/N \\
\hline
\hline 
15001387& 4119& 0.76&  2.59& $-$1.697& $-$0.210&    0.815&  0.392&  0.444&  0.452&     0.003& $-$0.190&  89  \\ 
15001907& 4184& 0.88&  2.34& $-$1.441&    0.259&    0.575&  0.407&  0.455&  0.180&     0.095& $-$0.088&  80  \\ 
24071427& 4067& 0.67&  2.41& $-$1.619&    0.328&    0.047&  0.414&  0.357&  0.386&  $-$0.098& $-$0.209&  98  \\ 
24100517& 4341& 1.17&  2.11& $-$1.533&    0.300& $-$0.074&  0.391&  0.257&  0.477&  $-$0.198& $-$0.129&  57  \\ 
24211267& 4126& 0.78&  1.22& $-$1.642& $-$0.124&    0.574&  0.144&  0.174&  0.229&  $-$0.102&    0.413&  86  \\ 
24237207& 4099& 0.73&  1.89& $-$1.778&    0.230&    0.325&  0.381&  0.427&  0.326&  $-$0.053&    0.035&  96  \\ 
38000367& 3970& 0.48&  2.95& $-$1.871&    0.478&    0.880&  0.478&  0.562&  0.193&     0.180& $-$0.398&  126 \\ 
38000407& 4014& 0.57&  1.73& $-$1.699&    0.415&    0.149&  0.445&  0.498&  0.256&     0.132& $-$0.143&  110 \\ 
38000567& 4025& 0.59&  1.58& $-$1.578&    0.334&    0.208&  0.401&  0.386&  0.275&     0.005&    0.143&  102 \\ 
38000737& 4093& 0.71&  1.98& $-$1.637& $-$0.423&    0.697&  0.230&  0.519&  0.592&  $-$0.100&    0.210&  99  \\ 
38001047& 4013& 0.56&  1.37& $-$1.478&    0.130&    0.707&  0.425&  0.376&  0.458&     0.065&    0.187&  101 \\ 
38001507& 4211& 0.94&  1.78& $-$1.419&    0.333& $-$0.018&  0.358&  0.418&  0.268&     0.095& $-$0.053&  69  \\ 
38001557& 4224& 0.96&  1.63& $-$1.644&    0.298&    0.356&  0.512&  0.486&  0.240&     0.120& $-$0.202&  78  \\ 
38002047& 4197& 0.90&  2.18& $-$1.274&    0.147&    0.542&  0.440&  0.367&  0.299&     0.059& $-$0.307&  78  \\ 
38002077& 4266& 1.04&  1.37& $-$1.534&    0.265& $-$0.021&  0.328&  0.414&  0.278&     0.043&    0.171&  69  \\ 
38002147& 4266& 1.04&  2.00& $-$1.357&    0.445& $-$0.138&  0.295&  0.337&  0.297&     0.003& $-$0.314&  66  \\ 
38002197& 4261& 1.02&  1.54& $-$1.373& $-$0.042&    0.587&  0.297&  0.371&  0.347&     0.034&    0.203&  66  \\ 
38002347& 4286& 1.07&  1.85& $-$1.294& $-$0.532&    0.692&  0.242&  0.197&  0.405&  $-$0.107&    0.091&  62  \\ 
38002827& 4305& 1.11&  1.48& $-$1.541&    0.100&    0.749&  0.401&  0.446&  0.411&     0.003&    0.392&  67  \\ 
38002877& 4324& 1.14&  1.63& $-$1.559& $-$0.047&    0.485&  0.422&  0.431&  0.437&  $-$0.110&    0.145&  63  \\ 
38002977& 4348& 1.19&  1.59& $-$1.338& $-$0.588&    0.861&  0.175&  0.552&  0.472&  $-$0.110&    0.494&  66  \\ 
38002987& 4341& 1.18&  1.84& $-$1.255& $-$0.640&    0.836& -0.487&  0.531&  0.503&  $-$0.105&    0.181&  59  \\ 
38002997& 4356& 1.21&  1.99& $-$1.507& $-$0.503&    0.561&  0.036&  0.450&  0.386&  $-$0.158&    0.208&  56  \\ 
38003047& 4353& 1.20&  2.21& $-$1.663&    0.083&    0.462&  0.448&  0.373&  0.392&  $-$0.124&    0.074&  62  \\ 
38003067& 4397& 1.19&  2.54& $-$1.671&    0.022&    0.659&  0.331&  0.308&  0.294&  $-$0.067& $-$0.193&  53  \\ 
38003117& 4363& 1.22&  2.13& $-$1.870&    0.094&    0.499&  0.408&  0.385&  0.352&  $-$0.062& $-$0.104&  61  \\ 
38003197& 4348& 1.19&  1.72& $-$1.460& $-$0.067&    0.521&  0.395&  0.401&  0.586&  $-$0.127&    0.183&  56  \\ 
38003237& 4385& 1.26&  2.30& $-$1.738&    0.222&    0.038&  0.440&  0.368&  0.335&     0.006& $-$0.007&  50  \\ 
38003657& 4362& 1.22&  2.21& $-$1.902&    0.299& $-$0.032&  0.378&  0.386&  0.415&  $-$0.061& $-$0.139&  59  \\ 
38003847& 4412& 1.31&  1.97& $-$1.242& $-$0.599&    0.570&  0.279&  0.467&  0.374&     0.043& $-$0.088&  45  \\ 
38004417& 4447& 1.38&  2.51& $-$1.726&    0.326&    0.508&  0.244&  0.252&  0.446&  $-$0.091& $-$0.086&  51  \\ 
38004437& 4445& 1.37&  2.12& $-$1.587&    0.180&    0.352&  0.413&  0.278&  0.487&  $-$0.102&    0.098&  47  \\ 
38004687& 4455& 1.39&  2.30& $-$1.743& $-$0.208&    0.552&       &  0.280&  0.385&     0.026&    0.178&  46  \\ 
38004707& 4407& 1.30&  2.16& $-$1.340& $-$0.157&    0.873&  0.421&  0.282&  0.384&     0.023& $-$0.087&  55  \\ 
38004717& 4435& 1.36&  1.79& $-$1.426&    0.426& $-$0.091&  0.491&  0.322&  0.269&     0.006& $-$0.127&  55  \\ 
38005967& 4274& 1.05&  1.88& $-$1.433& $-$0.545&    0.653&  0.170&  0.542&  0.340&  $-$0.018&    0.064&  68  \\ 
38007017& 4381& 1.25&  2.24& $-$1.179& $-$0.294&    0.926&  0.164&  0.534&  0.427&     0.031&    0.070&  56  \\ 
38009317& 4390& 1.28&  2.27& $-$1.793&    0.239&    0.204&  0.392&  0.432&  0.443&  $-$0.052&    0.049&  49  \\ 
38011977& 4428& 1.34&  1.88& $-$1.464&    0.332& $-$0.057&  0.599&  0.359&  0.351&     0.101&    0.124&  51  \\ 
41067037& 3953& 0.45&  2.47& $-$1.632&    0.040&    0.822&  0.105&  0.570&  0.363&     0.089& $-$0.269&  113 \\ 
38003167& 4337& 1.17&  1.94& $-$1.518&    0.261&    0.220&  0.418&  0.331&  0.316&     0.049&    0.154&  57  \\ 
38000597& 3946& 0.43&  1.45& $-$1.643&    0.298&    0.101&  0.449&  0.544&  0.134&     0.185& $-$0.084&  107 \\ 
\hline
\end{tabular}
\label{t:datam54}
\end{table*}

Following the procedure in Carretta (2021), we neglected the line at 7664.91~\AA\
(contaminated by strong telluric line) and we focused on the 7698.98~\AA\ line,
which is affected by a weaker line. This spectral region was then cleaned
 by dividing for a synthetic spectrum of the telluric lines over the region
7681-7710~\AA, following the procedure described for the [O~{\sc i}] forbidden
line in Carretta et al. (2006).

On these cleaned spectra, we used the package ROSA (Gratton 1988) to measure the
equivalent widths (EWs) of the K line. For the analysis we adopted the same 
atmospheric parameters derived by Carretta et al. (2010c). As widely discussed
(see Carretta et al. 2014, Carretta 2021, Mucciarelli et al. 2015), when using
$v_t$ values derived by moderately weak Fe lines together with strong lines as
those of Ba~{\sc ii} or K~{\sc i}, a trend of abundance as a function of $v_t$
may occur. A classical way to bypass the problem is to calibrate $v_t$ as a
function of a parameter such as the surface gravity, adopting a constant
metallicity for all the cluster stars. However, this is not the best procedure
for a GC showing an intrinsic metallicity spread (about 0.19 dex) such as M~54. The
resulting abundances then show an unavoidable trend, decreasing as $v_t$
increases. This effect does not changes the results, owing to the small internal
error estimated for $v_t$ (Carretta et al. 2010c) with respect to the large
spread we found in K (see below).

For strict homogeneity with previous works (Carretta et al. 2013a, Mucciarelli
et al. 2015, 2017, Carretta 2021), we corrected K abundances  for departure from
LTE using a multivariate interpolation as a function of temperature, gravity,
metallicity, and equivalent width of the K~{\sc i} line from the set of models by
Takeda et al. 
(2002)\footnote{$http://www2.nao.ac.jp/\sim takedayi/potassium\_nonlte/7699/$}.
Reggiani et al. (2019) produced a much more extensive grid of
NLTE corrections, based on more updated atom model and collisions  with hydrogen
and electrons. We downloaded their grid (updated 16 April 2021) from CDS and
computed the corrections with the same interpolatory function as above. We found
that an almost rigid shift by 0.186 dex ($rms=0.040$ dex, 42 stars) in [K/Fe]
results from differences in the corrected abundances, with the final [K/Fe]
ratios being higher when using the grid by Reggiani and collaborators. While
absolute values differ, ranking is well preserved and we keep to the previous
corrections for homogeneity.

We derived sensitivities of K abundances to changes in the atmospheric
parameters. Coupled to the sensitivities in Fe and to the internal uncertainties
in each parameters as estimated in Carretta et al. (2010c), we obtained star to
star errors. These were summed in quadrature to uncertainties in the $EW$
measurements, estimated by Cayrel's (1988) formula. In the ranges S/N$>100$,
60$<$S/N$<$100, and  S/N$<60,$ the total internal error budget amounts to 0.081
dex, 0.089 dex, and 0.103 dex, respectively. We then adopted the average 0.091 dex
as the typical internal error associated with the [K/Fe] ratios in M~54.

The derived [K/Fe] ratios, corrected for non-LTE effects, are listed in the
second to last column of Table~\ref{t:datam54}. We adopted the solar abundance
of K from Anders and Grevesse (1989).

\section{Results}

\begin{figure*}
\centering
\includegraphics[scale=0.30]{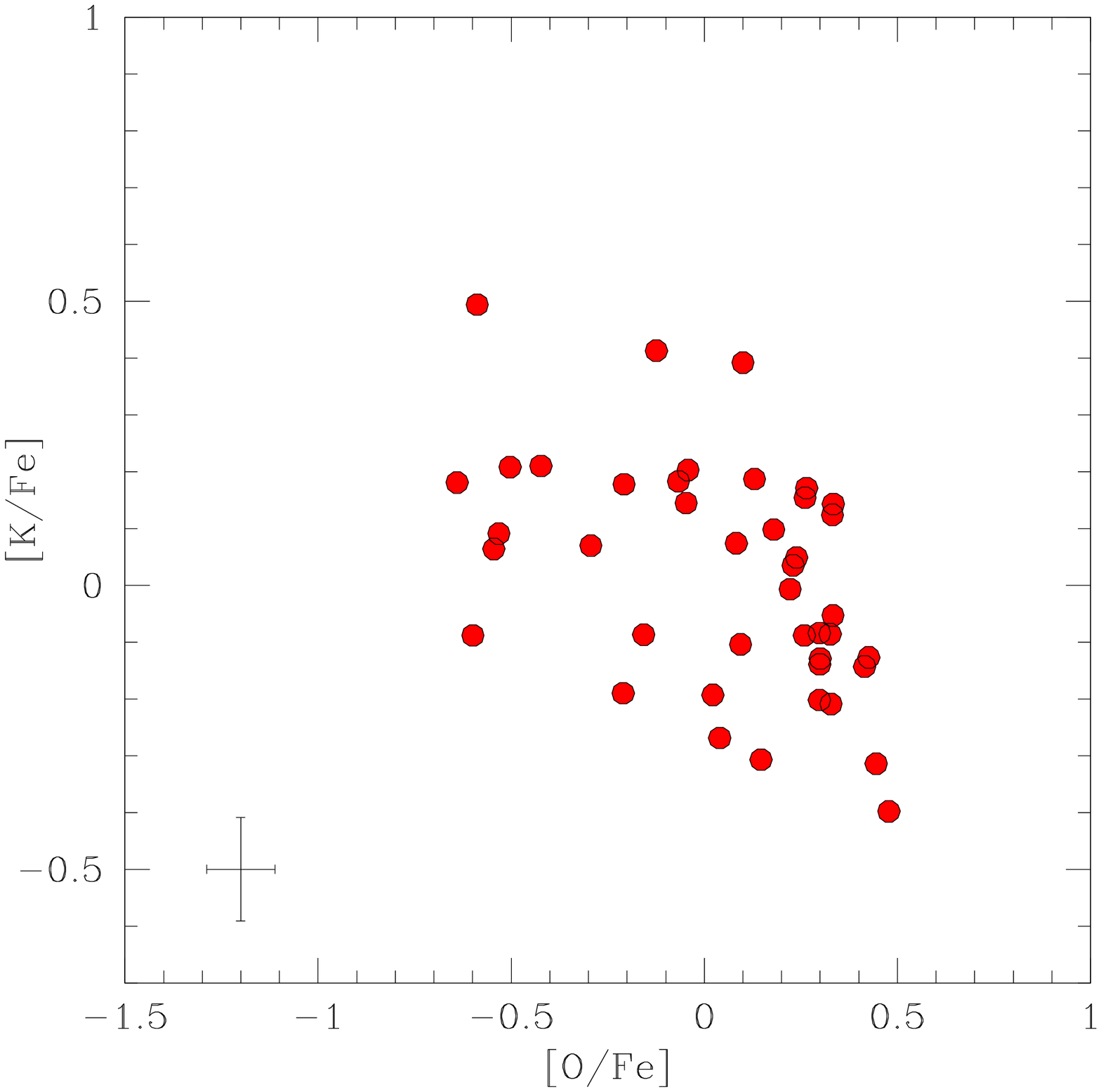}\includegraphics[scale=0.30]{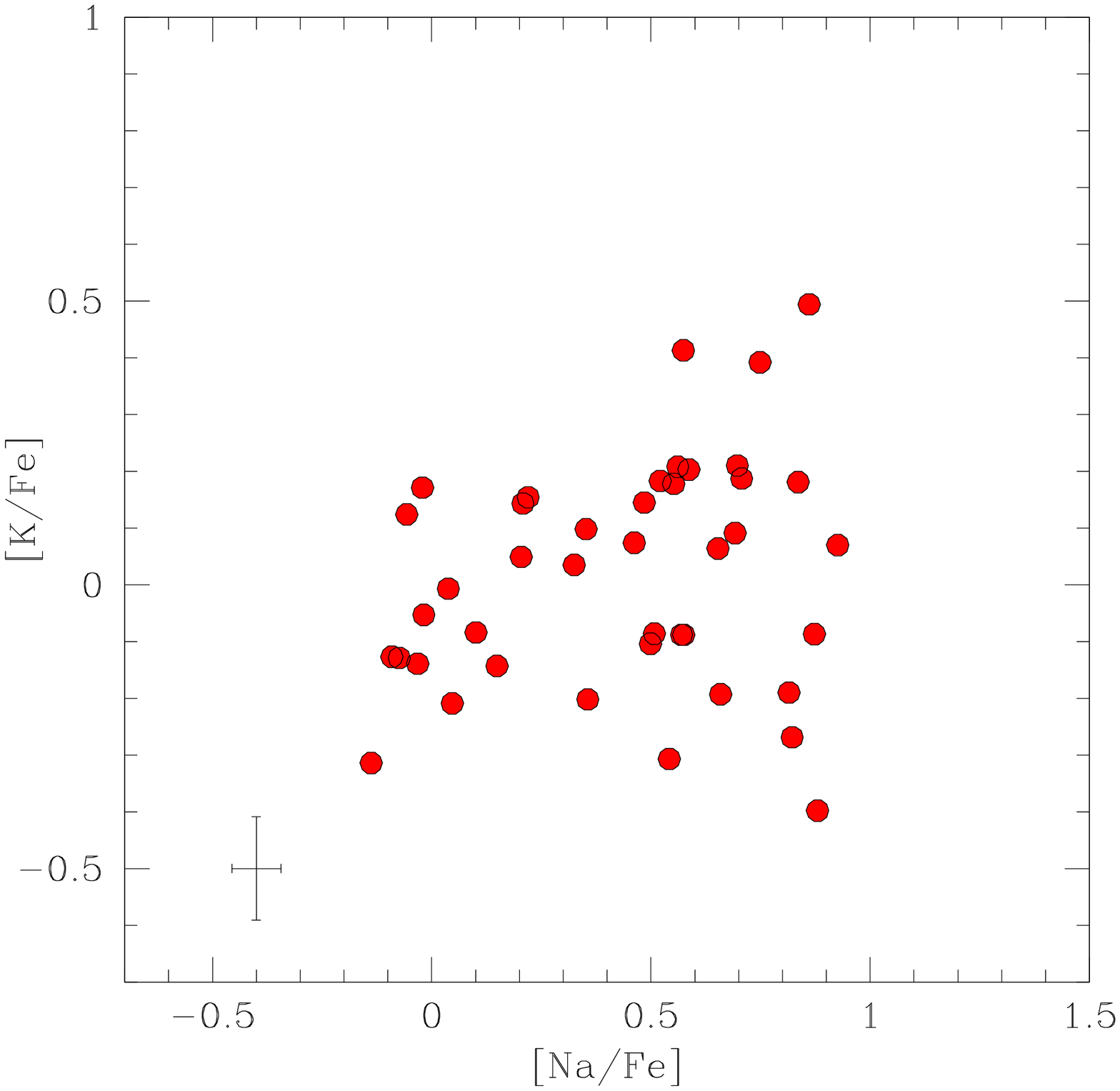}
\includegraphics[scale=0.30]{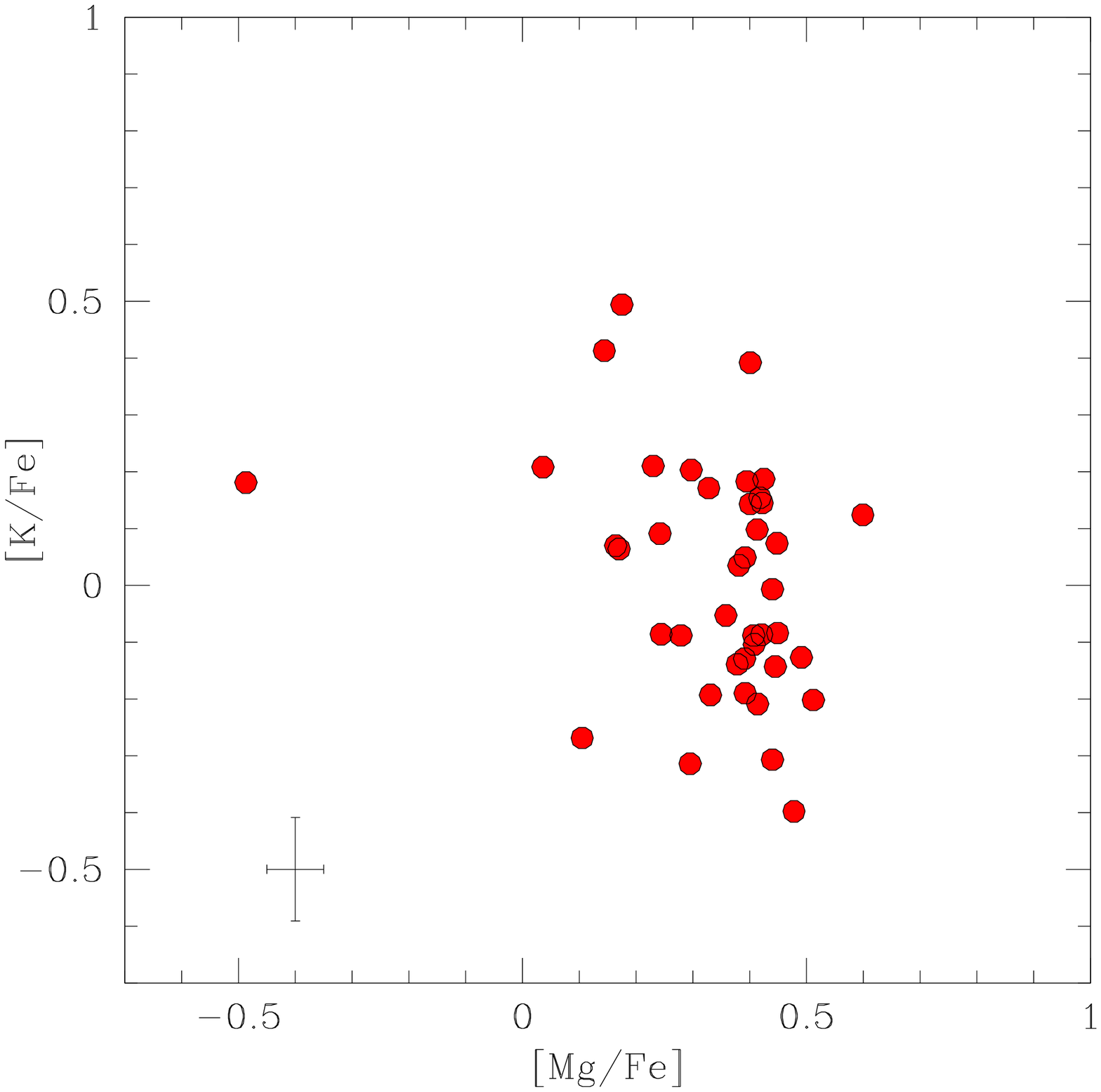}\includegraphics[scale=0.30]{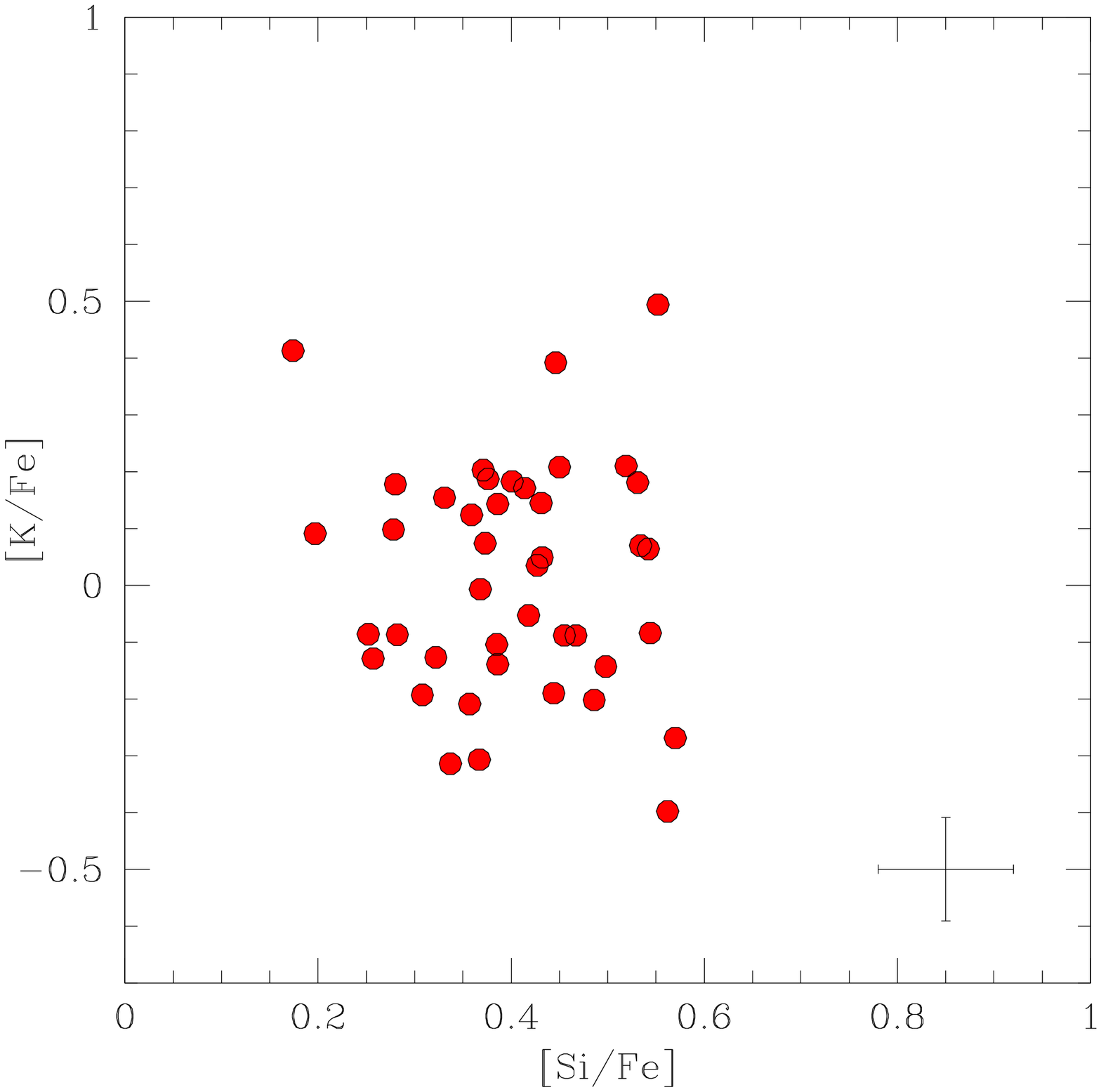}
\includegraphics[scale=0.30]{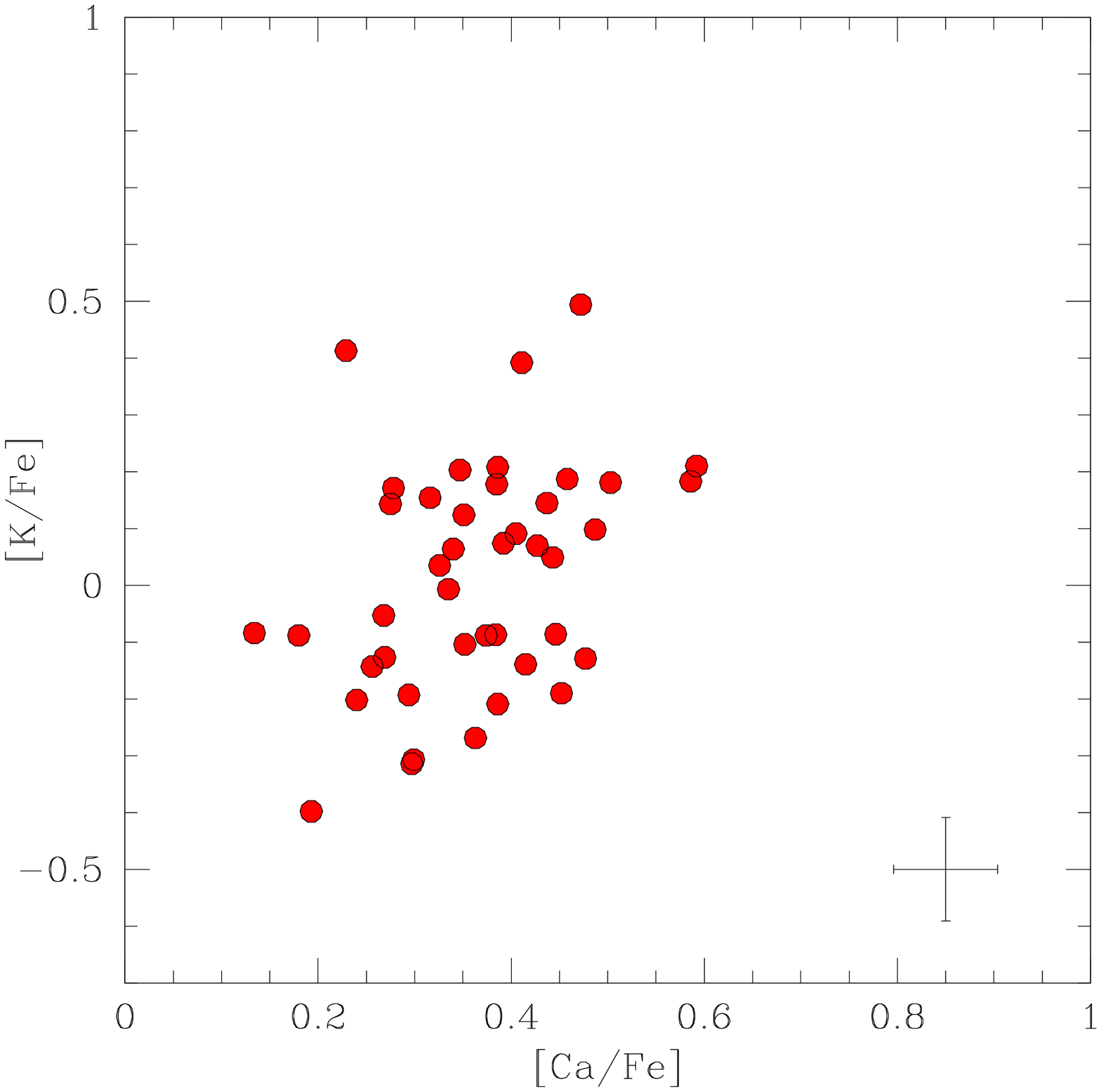}\includegraphics[scale=0.30]{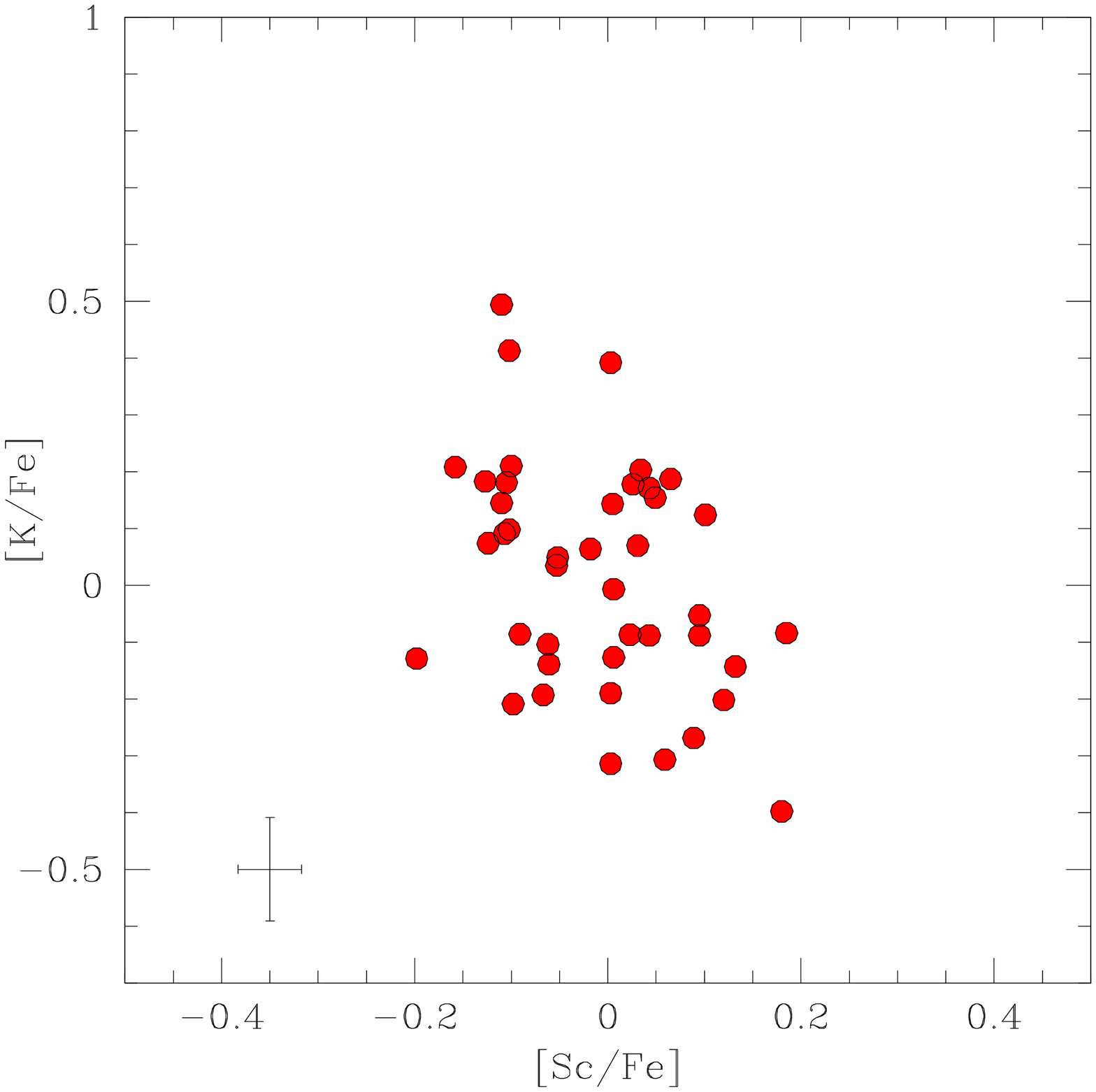}
\caption{Abundance ratios of [K/Fe] in M~54 as a function of abundances of O, Na,
Mg, Si, Ca, and Sc. Internal, star to star errors from the present study and
from Carretta et al. (2010c) are shown in each panel.}
\label{f:figtot}
\end{figure*}

In M~54 we observe a large spread of K abundances. The rms about a mean value
[K/Fe]=0.015 dex is 0.202 dex (42 stars), the interquartile range is
IQR[K/Fe]=0.295, and the [K/Fe] values in this GC extends over almost 1 dex,
with no trend as a function of effective temperature.
Results for K are summarised in Fig.~\ref{f:figtot}, where the [K/Fe] ratios are
compared to other elements involved in the proton-capture reactions occurring in
various temperature regimes: O, Na, Mg, Si, Ca, and Sc (Carretta et al. 2010c).

The panels on the left side of this figure show the most significant relations.
Potassium abundances are anti-correlated with elements that are depleted in multiple
stellar populations, such as O and Mg. The anti-correlation [K/Fe]-[O/Fe] is highly
significant, the Pearson correlation coefficient being $r=-$0.471 with 42 pairs.
A Student's test for the null hypothesis that the true correlation is zero 
returns a probability $p=1.6\times 10^{-3}$. 
This means that the large spread in K observed in M~54 is likely due to the same
network of proton-capture reactions acting at early times to self-enrich the
intra-cluster medium with product of hot H-burning, for instance matter severely depleted
in O.

Although there is no doubt that the highest K abundances are detected in stars
with the lowest Mg values (central left panel in Fig.~\ref{f:figtot}),
statistically the anti-correlation between K and Mg abundances is only
marginally significant. Formally we obtained $p=0.055$, slightly exceeding the
typical threshold value of 0.05 .
The explanation is twofold. First, a straight line may
be not the best functional form to approximate the K-Mg relation. Second,
although we observe a spread in K abundances of about 0.9 dex, in M~54 there
is an intrinsic paucity of very Mg-poor stars, those with which  the highest [K/Fe]
values should be associated. In our sample of 42 stars, only one has
[Mg/Fe]$<0$ (2.4\%). There is not a selection bias affecting our present sample:
among the 76 giants examined by Carretta et al. (2010c) we only found two stars
with [Mg/Fe] below zero (2.6\%).

The lack of stars with severe depletion in Mg seems to be an actual feature of
M~54, at odds with what was found in NGC~2808 (Carretta 2015) and NGC~4833 (Carretta
et al. 2014), despite M~54 being a more massive GC. We also note that the
anti-correlation K-Mg is found regardless of the metallicity, being traced in
both the metal-rich and metal-poor components individuated in M~54 (Carretta et
al. 2010c).

Other relations, in particular with elements expected to be enhanced by
proton-capture reactions, appear to be less significant (see right side panels in
Fig.~\ref{f:figtot}). A correlation with Na appears not to be significant,
although expected, owing to the neat anti-correlation K-O and the quite extended
Na-O anti-correlation found in M~54 (Carretta et al. 2010c). There seems to be
no relation between K and Si. We recall that in M~54, a clear Si-Al correlation
and Mg-Al anti-correlation were found, although evidence of a Mg-Si
anti-correlation is weak (Carretta et al. 2010c).
We note that Al abundances were derived in Carretta et al. (2010c) only for
six stars with UVES spectra and only one star (38003167) is included in the
present sample with derived K abundances. This occurrence prevents us from
drawing any further conclusion on the Mg-Al cycle, in particular on the peculiar
interplay found by Masseron et al. (2019) in very metal-poor GCs such as M~15 and
M~92, where the most Mg-depleted stars are not always accompanied by the most
enhanced Al abundances. Clearly, more Al data would be welcome for M~54.

Finally, it is interesting to note that a significant K-Ca correlation (bottom
left panel in Fig.~\ref{f:figtot}) does exist in M~54 ($p=1.4\times 10^{-2}$),
whereas an anti-correlation with Sc is apparent.  However, in the extensive
census of 77 GCs, Carretta and Bragaglia (2021) found excesses for both Ca and
Sc in M~54 with respect to an unpolluted sample of field stars that show only
the effects of nucleosynthesis by supernovae. 

\begin{figure}
\centering
\includegraphics[scale=0.40]{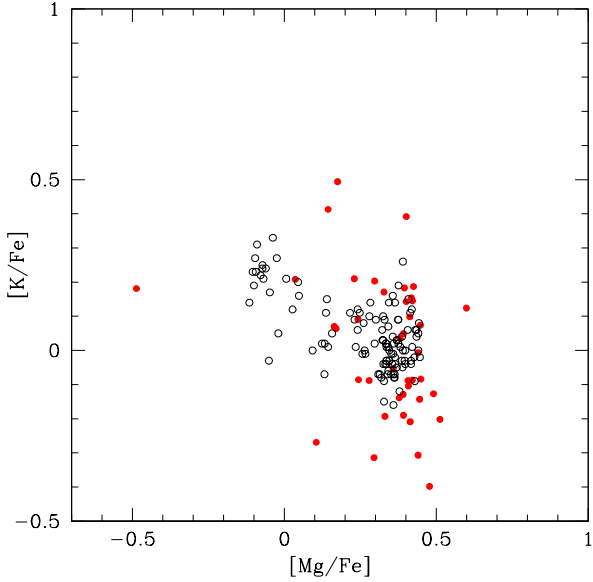}
\caption{Comparison of the K-Mg anti-correlation in M~54 (present work, filled
red points) to the one in NGC~2808 (Carretta 2015, Mucciarelli et al. 2015, open
circles).}
\label{f:figkmg28}
\end{figure}

The dearth of Mg-poor stars in M~54 is clearly seen by comparing our results
with those in NGC~2808 (Fig.~\ref{f:figkmg28}). While the entire spread in K 
is about half of the one obtained in M~54, NGC~2808 hosts a sizeable
population of Mg-poor stars. As a consequence, the K-Mg relation is steeper in
M~54, accounting for the marginally significant statistics, as seen above.

\begin{figure}
\centering
\includegraphics[scale=0.40]{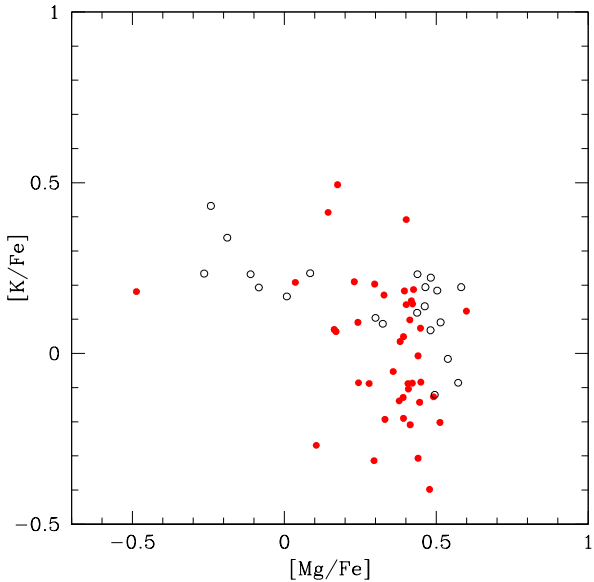}
\caption{Same as for Fig. 4, but for $\omega$ Cen using infrared APOGEE
data from M\'esz\'aros et al. (2020, empty circles).}
\label{f:figkmgocen}
\end{figure}

Unfortunately a similar comparison with optical data for $\omega$ Cen is
hampered by the unexplained trend in temperatures of K abundances derived by
AG22. The comparison with the 21 stars in M\'esz\'aros et al. (2020) passing
their quality cuts is shown in Fig.~\ref{f:figkmgocen}. With this sample from APOGEE data, the comparison is similar to the one for
NGC~2808: a larger spread in K abundances for M~54 accompanied by the presence
of Mg poor stars in $\omega$ Cen. The offset in [K/Fe] is probably due to the
NLTE corrections: M\'esz\'aros et al. (2020) do not mention any, whereas the 
corrections made with the grid of Takeda et al. (2002) lead to NLTE K abundances
that are lower by about -0.5 dex.
M\'esz\'aros et al. claim a weak K-Mg anti-correlation in $\omega$ Cen, but we
found that despite the limited size of their sample, the anti-correlation is
significant ($p=1.1\times 10^{-3}$)\footnote{We note that above their Fig. 10, M\'esz\'aros et al. (2020) wrote [Fe/H]$<-1.5$ among their
quality cuts. Their figure for $\omega$ Cen is, however, reproduced only if
[Fe/H]$>-1.5$ is used.}. Hence, the existence of a K-Mg anti-correlation also in
$\omega$ Cen seems to be well assessed, despite possible problems in the
analysis (AG22) or the weakness of K lines at low metallicity in infrared data
(M\'esz\'aros et al. 2020).

\section{Summary}

We presented the first study of K abundances in NGC~6715 (M~54), the second most
massive GC in the Milky Way and a former member of the Sagittarius dwarf 
galaxy. We analysed FLAMES-GIRAFFE archival spectra with the high-resolution
setup HR18 of 42 RGB stars in this GC.

The [K/Fe] ratios, obtained from the K~{\sc i} 7698.98~\AA\ line and corrected
for NLTE effects with the grid by Takeda et al. (2002) span a range of about a
full dex. The [K/Fe] and [O/Fe] ratios in M~54 are found to be anti-correlated 
to a high level of significance, indicating that the large excess in K is likely
due to the same self-enrichment process that generates the multiple population
phenomenon in GCs. A marginally significant K-Mg anti-correlation is also
observed. On the other hand, the expected correlations with species that are
expected to be enhanced in proton-capture reactions is less clear, apart from a
significant K-Ca correlation, confirming the Ca excesses with respect to
unpolluted field stars recently detected by Carretta and Bragaglia (2021) in
this GC.
The comparison with NGC~2808 shows that the observed pattern is well explained
by the large spread of K in M~54 accompanied by a lack of very Mg-poor stars,
which are instead hosted in the first GC as well as in $\omega$ Cen.

Our data confirm the predictions by Carretta et al. (2014) based on the
diagnostic plot [Ca/Mg] versus [Ca/H], independently using K abundances  as 
tracers of the action of hot H-burning at very high-temperature regime. This
regime is predicted (see Ventura et al. 2012, Prantzos et al. 2017) to be  more
efficient in massive GCs populating the low-metallicity tail of the [Fe/H]
distribution of GCs. Carretta (2021) showed that the few GCs where significant
variations in K abundances are detected so far (NGC~2419, NGC~2808, NGC~5139 and
NGC~4833) tend to populate the high-mass (represented by the cluster total 
absolute magnitude) and low-metallicity region of the $M_V$-[Fe/H] plane. To
these GCs, we now also add M~54. In the same region are also located most of the
GCs where excesses in Ca and or Sc were detected with respect to unpolluted
field stars by Carretta and Bragaglia (2021).
All these observations concur to outline a picture where particular conditions
may allow the proton-capture reactions to bypass, in part, the usual production 
of Al, reaching up to heavier species such as K, and also possibly Ca and Sc. 

Since the interplay between polluters acting at different temperature (mass)
regimes may offer precious constraints to the theoretical models, it is important
to gather as much observational material as possible of different species
involved in proton-capture reactions.
In this sense, the present study represents another step towards the nature
of the still elusive FG polluters.

The primary dependence of multiple populations on the total mass of GCs has been well
known since the first studies of the Na-O and Mg-Al anti-correlations with
respect to global cluster parameters (Carretta 2006, his
figure 12). It is thus easy to predict large variations in light elements in the
two most massive GCs in the Galaxy, even for species requiring high temperatures
for nuclear burning. For M~54, data are presented in the present work. 
Concerning $\omega$ Cen, we note that M\'esz\'aros et al. (2020) already
provided convincing evidence of large K variations, using only stars more
metal-rich than [Fe/H]$=-1.5$ dex. Unfortunately, the K lines in the H band become
very weak at low metallicities, a regime where the production of K from 
proton-capture reactions is predicted to be more efficient.
The analysis of optical spectra looks to be more promising, once spurious trends of K 
as a function of the effective temperature of stars are able to be corrected.

\begin{acknowledgements}
This paper is based on data obtained from the ESO Science Archive Facility under
request number 593222.
I wish to thank Yazan Al Momany for providing the original WFI photometry for
M~54 and Angela Bragaglia for valuable help and useful discussions.
This research has made large use of the SIMBAD database (in
particular  Vizier), operated at CDS, Strasbourg, France, and of the NASA's
Astrophysical Data System.
\end{acknowledgements}

\end{document}